\documentclass[prb,aps,twocolumn,showpacs,preprintnumbers,amsmath,amssymb]{revtex4}
\usepackage{epsf}

\begin{document}
\draft
\newcommand{\ve}[1]{\boldsymbol{#1}}

\title{Hydrogen transport in superionic system Rb$_3$H(SeO$_4$)$_2$: a revised
cooperative migration mechanism}
\author{N.~Pavlenko$^{\rm 1}$,
A.~Pietraszko$^{\rm 2}$, A.~Pawlowski$^{\rm 3}$, M.~Polomska$^{\rm 3}$, I.V.~Stasyuk$^{\rm 1}$, and
B.~Hilczer$^{\rm 3}$}

\address{
$^1$ Institute for Consensed Matter Physics, 79011 Lviv, Ukraine\\
$^2$ Institute of Low Temperatures and Structural Research PAN, Wroclaw, Poland\\
$^3$ Institute of Molecular Physics PAN, Smoluchowskiego 17, 70169 Poznan, Poland
}


\begin{abstract}
We performed density functional studies of electronic properties and mechanisms of hydrogen transport in
Rb$_3$H(SeO$_4$)$_2$ crystal which represents
technologically promising class M$_3$H(XO$_4$)$_2$ of proton conductors (M=Rb,Cs, NH4; X=S,Se). 
The electronic structure calculations show a decisive role of lattice dynamics in the
process of proton migration. In the obtained revised mechanism of proton transport, the 
strong displacements of the vertex oxygens play a key role in the establishing the continuous
hydrogen transport and in the achieving low activation energies of proton conduction
which is in contrast to the standard two-stage Grotthuss mechanism
of proton transport. Consequently, any realistic model description of proton transport
should inevitably involve the interactions with the sublattice of the XO$_4$ groups.
\end{abstract}

\maketitle

\section{Introduction}

Last years demonstrate a continuous increase of research activity in the field of hydrogen conductors.
This can be explained by great technological perspectives of hydrogen conducting materials
for applications in solid-state hydrogen fuel cells, hydrogen storage
and electrochemical devices \cite{norby,haile,nature1,dft1}.
A central problem in 
fuel-cell and hydrogen batteries technology is
the development of cheap and efficient materials for electrochemical elements which can be used
for the chemical storage and transformation of chemical to electrical energy. In the functional
properties of hydrogen devices, the transport of hydrogen through the conducting part to cathode
plays a key role which makes the proton transport mechanisms one of the central problems of current
experimental and theoretical studies \cite{merinov,bjorketun,zhang}.

In the wide range of proton conducting materials, the crystals with superionic phases
M$_3$H(XO$_4$)$_2$ (M=Rb,Cs, NH$_4$; X=S,Se) are of especial
interest due to their high proton conductivity coefficients which in the high temperature disordered
state can reach values of the order $10^{-2}-10^{-1}$~$\Omega^{-1} cm^{-1}$. In these phases,
the quasi-two-dimensional proton transport is characterized by a dynamically disordered
network of virtual hydrogen bonds dynamically establishing between the oxygens O(2) of nearest groups XO$_4$. Although
on the first sight the transport of protons in hexagonal (001) planes seems to be quite a strainforward
process explained in terms of the two-stage Grotthuss mechanism \cite{belushkin,lechner,yamada}, 
the real mechanism of proton migration
is still not sufficiently well examined and understood. The reason for this difficulty is a complex cooperative character of proton
migration which should strongly interact with the dynamics of the ionic environment, the fact supported
by numerous experimental Raman, Infrared, x-ray and NMR spectroscopy studies \cite{merinov1,baran2,merinov2,baranov,
bohn,dolinsek,hilczer}.

In the theoretical works \cite{dolinsek,plakida,kamimura,stasyuk,pavlenko,pavlenko2}, the 
two-stage Grotthuss mechanism
is frequently considered within phenomenological free-particle approches which
typically contain two stages: (1) translation
of proton within a hydrogen bond between the nearest ionic groups XO$_4$; (2) inter-bond proton
transfer related to reorientations of the ionic groups XO$_4$.
Howerer, the experimental studies demonstrate that 
the transport of protons is a highly complex process which can be characterized as a transfer 
of hydrogen ion in the course of the reactions of creation and breaking of HXO$_4$ bonds.
As a consequence of such reacting environment with dynamically redistributing
electronic density, the protons in conducting materials
cannot be described as almost free particles.

Despite the existing theoretical studies of the low-temperature proton ordered phases and
proton conduction in M$_3$H(XO$_4$)$_2$ by phenomenological models \cite{plakida,kamimura,stasyuk,pavlenko,pavlenko2},
up to date there is no first-principle-investigations of these systems. The recent first-principle molecular dynamics
studies of superprotonic phase transition in a related
superionic system CsHSO$_4$ \cite{chisholm} were focused on the structural 
transformations and reorientational motion ofHSO$_4^-$ groups, without the studies of the proton migration mechanism. 
Another {\it ab-initio} study of the proton conduction in CsHSO$_4$ has beed devoted to the estimation of diffusion barriers,
without a detailed consideration of the role of oxygen network in the proton transport mechanism \cite{ke}.   
The lack of the first-principle investigations of the transport mechanism in M$_3$H(XO$_4$)$_2$ can be explained
by highly complicated crystal structure of these systems involving several structural transformations
which makes {\it ab-initio} calculations extremely time-consuming and challenging. 
Our present studies of the hydrogen transport are based on the
density functional theory (DFT) with the use of the linearized augmented plane wave
method (LAPW) implemented in the WIEN2k package \cite{wien2k}. The
LDA (Local Density with nonlinear core corrections) and GGA (Generalized Gradient) approximations of the DFT developed by
Perdew, Burke and Ernzerhof \cite{pbe}, and Perdew and Zunger \cite{perdew_zunger}
have been employed in our calculations.
In the GGA, the additional gradient terms of the electron density are added to the exchange-correlation energy
and its corresponding potential, which is beyond the local density approximation (LDA) of the DFT.
In our studies of the proton migration paths, we also employ a nudged-elastic-band method within the 
pseudopotential approach implemented in the Quantum Espresso package \cite{qe} which allows to perform
the dynamical relaxation of the atomic surrounding along the steps of the proton conduction processes.

As an attempt to account for the coupling of mobile protons with the
dynamical distortions of the groups XO$_4$, the existing microscopic description of the proton transport 
considers the interaction of protons with the displacements of the vertex 
oxygens O(2) of XO$_4$ involved in the hydrogen bonding\cite{pavlenko}. Such a coupling is
presented in terms of the interaction of proton charge with the optical phonon modes
corresponding to the anti-phase stretching vibrations of the O(2)-ions and is classified in terms
of ``protonic polaron'' concept. The calculated in this approach coefficients of the proton conductivity
are consistent with the experimental measurements performed in the superionic phases \cite{pawlowski}. 
However, the polaronic concept introduced in Ref.~\onlinecite{pavlenko} 
involves a number of phenomenological parameters like polaron binding energy which have been
estimated by the fitting of the calculated conductivity to the experimetally measured values. To 
verify the role of the dynamical distortions in the proton migration mechanism, in this work we present
a detailed DFT study of the mechanism of hydrogen migration in
the system Rb$_3$H(SeO$_4$)$_2$, which belongs to the M$_3$H(XO$_4$)$_2$-class.

To investigate the role of the proton environment, two
aspects of the hydrogen transport have been analyzed. The first aspect is related to the structure
and positioning of hydrogen-bonded oxygen ions O(2) in the dynamical network. The present DFT calculations 
demonstrate the strong displacements of O(2) from the high-symmetry positions on the three-fild axis
in the equilibrium state, which allows to verify the concepts of the O(2)-structural positioning discussed in
the previous x-ray and NMR studies \cite{baranov,melzer}.
As the second aspect, we have analyzed the migration paths of protons in the configuration
of instantaneously relaxed atoms. The performed DFT-calculations allow to derive the energy profiles
along the proton migration paths and to investigate the displacements of each atom involved in the
proton surrounding. As a result, the present studies give a possibility to obtain the first-principle
estimates of the hydrogen-bond lengths and energetic parameters involved in the phenomenological modeling
and to verify the concept of protonic
polaron introduced in the previous phenomenological description of proton transport. 
Our findings also allow to improve several experimental conclusions obtained in the 
previous x-ray and NMR experiments with Rb$_3$H(SeO$_4$)$_2$ \cite{baranov,melzer}, in particular to clalify the 
positioning
of the vertex oxygens O(2) during the proton migration and to examine the proton positions 
on the hydrogen bond at the intrabond proton transfer. As a result, the present DFT results
demonstrate a central role of oxygen displacements for
the migration of hydrogen and emphasize in this way 
a key role of the proton environment for the establishing the proton transport.

The rest of the paper is organized in three sections. In Section~\ref{methods}, the theoretical methods employed
in our studies are discussed. Section~\ref{results} presents the results of the electronic structure
calculations and of the calculations of the proton migration paths and oxygen positioning. 
In Section~\ref{conclusions}, we summarize the main results and conclusions of our work.

\section{THEORETICAL MODELS AND METHODS} \label{methods}

\subsection{Structural concepts of hydrogen-bonded network}

\begin{figure}[ht]
\epsfxsize=9.5cm \centerline{\epsffile{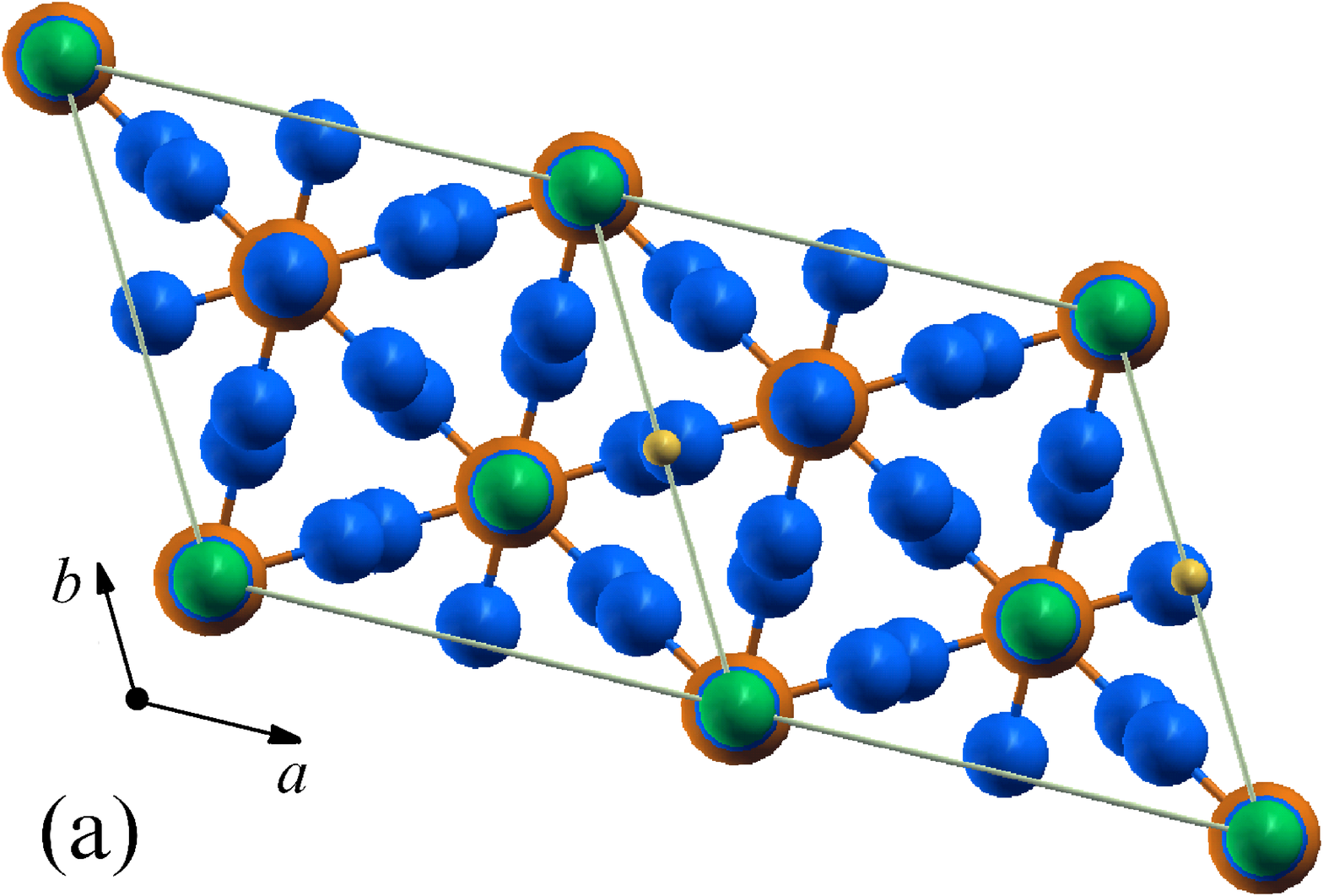}}
\epsfxsize=6.5cm \centerline{\epsffile{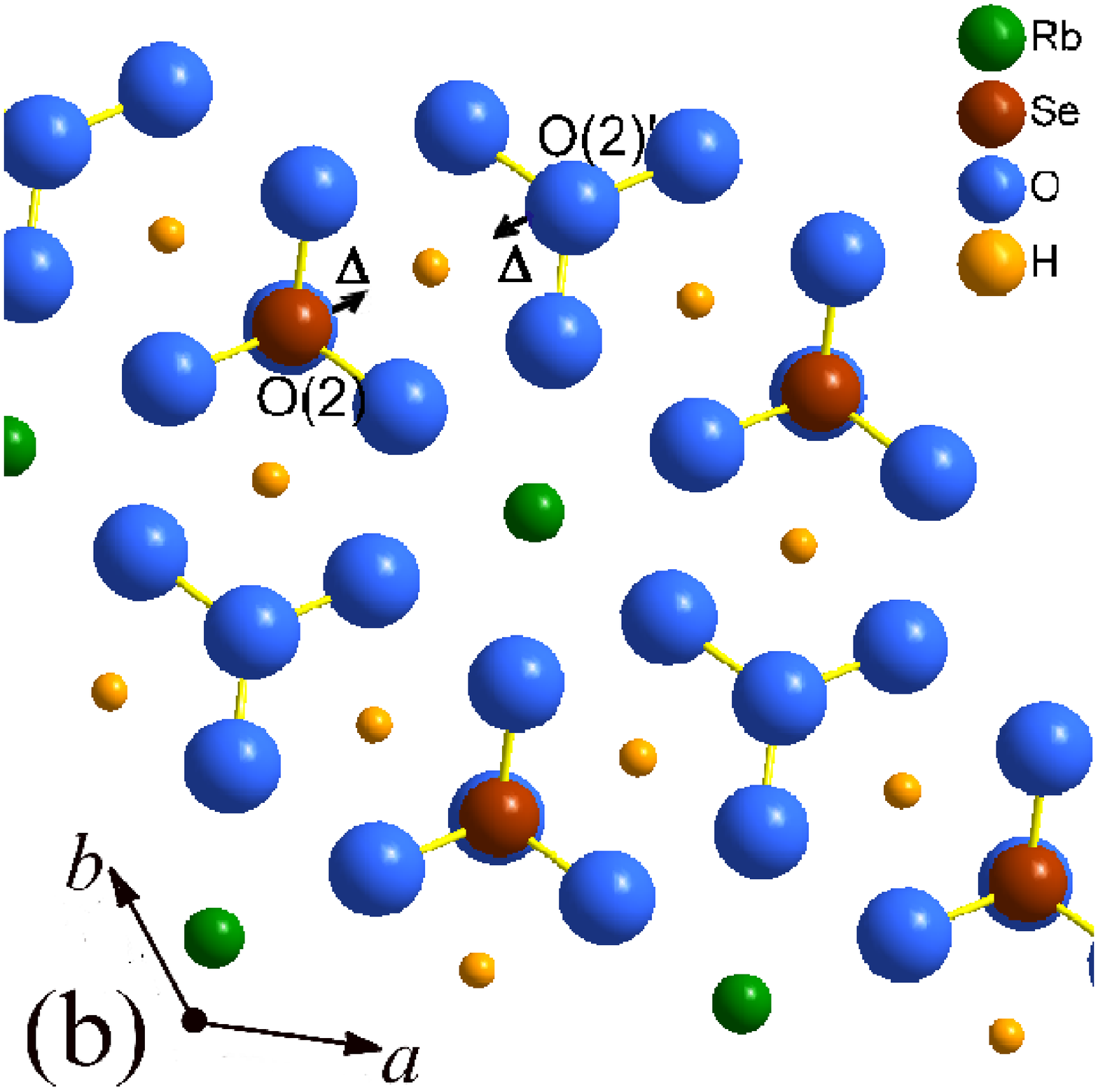}} \caption{
(a) Schematical presentation of the crystal structure of Rb$_3$H(SeO$_4$)$_2$
in the superionic phase in projection on the hexagonal (001) plane, where the rhombohedral unit cell
of R$\bar{3}$m symmetry is doubled along the (100)-direction. One of the virtual hydrogen bonds with the proton
centered on the bond is shown for simplicity. 
Here the small orange circles denote the protons, the oxygen atoms are indicated by
blue, Rb atoms by green and Se by brown circles.
(b) Atomic configuration of hydrogen-bonded network of SeO$_4$ tetrahedra in superionic phase
in the proton conducting layer corresponding to a slab ($-0.13c\le z \le 0.13c$) in the hexagonal (x,y) plane,
with the lattice constant $c=22.629$\AA.
The orange circles indicate the possible positions for the disordered protons occupied with a probabibity $1/3$.
} \label{fig1}
\end{figure}

In the high-temperature disordered phase of Rb$_3$H(SeO$_4$)$_2$ schematically shown
in Fig.~\ref{fig1}(a), hydrogen transport
occurs in a network of ``virtual'' (equally probable) hydrogen bonds which are occupied by
a proton with equal probability $1/3$. This network (see Fig.~\ref{fig1}(b) for details) 
is established between the vertex oxygens O(2) of the tetrahedra SeO$_4$ which can be chemically bonded
to the three nearest tetrahedra by hydrogen bonds.

\begin{figure}[ht]
\epsfxsize=8.5cm \centerline{\epsffile{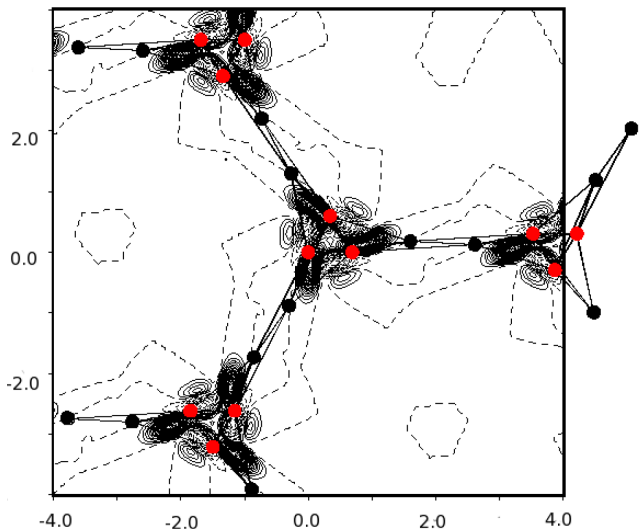}} \caption{
Electron charge density maps of the nearest neighbouring SeO$_4$ groups of Rb$_3$H(SeO$_4$)$_2$ in
the disordered superionic state at $T=456$~K. Here the red dots show the possible positions of the
vertex oxygens O(2) around the three-fold axis and the black dots correspond to the
location of disordered protons on O(2)-H-O(2)' hydrogen bonds.
} \label{fig2}
\end{figure}

The experimentally deduced structure of the network of 
hydrogen-bonded tetrahedra at $T=456$~K projected on (001) plane
is represented by the electronic difference density map
in  Fig.~\ref{fig2}. This map has been calculated from the x-ray diffraction
data using SHELXL-93 programs. The central feature detected in Fig.~\ref{fig2} is a disorder
of the hydrogen-bonded vertex oxygens O(2) (shown by red dots) between three structurally equivalent
positions. The hydrogen bonds are schematically presented by the solid lines where
the double black dots denote the positions for proton in a double-well potential.

In the earlier work by Baranov et al. \cite{baranov}, a different concept of the
structure of hydrogen-bonded network is proposed for the analysis of the
x-ray diffraction data. In this concept, the oxygens O(2) are statically located
in high-symmetry positions on three-fold axis.
which also implies a static character of the 
of SeO$_4$ tetrahedra and leads to a possibility of the movement
of free protons in the disordered superionic phase. Therefore, the arrangement
of the vertex oxygens O(2) is still an open question which 
has a prime importance for the understanding the nature of proton
transfer mechanism in these systems.

In the first concept of the structural
arrangement of O(2) in the superionic phase of Rb$_3$H(SeO$_4$)$_2$
signified in Ref.~\onlinecite{baranov}, the oxygens O(2) are located in the
high-symmetry positions on the three-fold axis and form weak hydrogen
bonds with the length of the covalent bond O(2)-H approaching $1.76$~\AA. 
In distinction from the high-symmetry case,
the second type of oxygen arrangement discussed in Ref.~\onlinecite{baranov,hilczer} assumes the
distortions of O(2) from the three-fold axis towards the hydrogen at the formation of the O(2)-H-O(2)'
bond. To prevent a break of high symmetry in the disordered superionic phase caused by such distortions,
a dynamical disorder of each O(2) between three equally probable positions
is assumed. These positions are related to each other by a rotation by $120\deg$ around
the three-fold axis (see Fig.~\ref{fig2} where the three equivalent positions of distorted O(2)
are marked by red dots)).

To analyze the equilibrium positions of the oxygens O(2), we performed the GGA calculations
of the total energy of the system
for the different values of the displacements $\Delta$ of the oxygen O(2) from the positions of the
three-fold axis towards a migrating proton, a process which is 
schematically shown in Fig.~\ref{fig1}(b).
In the calculations of the electronic structure of Rb$_3$H(SeO$_4$)$_2$,
the lattice constants of the unit cell of the trigonal R$\bar{3}$m symmetry in
hexagonal coordinates were fixed to the values obtained
from the structural analysis \cite{baranov}: $a=6.118~\AA$ and $c=22.629~\AA$.
Technical details include a GGA method developed by
Perdew, Burke and Ernzerhof \cite{pbe} on a $4\times 2 \times 3$ $\ve{k}$-point grid
representing a mesh of 40~$\ve{k}$-points in the first Brillouin zone. The comparison of
the total energy calculated for the mesh of 80~$\ve{k}$-points gives the difference
between the total energies $E_{tot}(80)-E_{tot}(40)=0.00136$~eV which allows us to
use the 40~$\ve{k}$-point mesh for the GGA-calculations.

The full-potential method \cite{wien2k} employs an expansion of the electronic potential inside the atomic spheres 
via a full number of the electronically occupied local orbitals 
(Rb, Se, O and H atoms in our case),
whereas the potential outside the spheres is constructed as a plane-wave expansion
with a plane-wave cutoff given by 667~eV. 
Such a method allows to achieve high accuracy in the description of the electronic structure, 
total energy and electronic density of states.
The dynamical disorder in the superionic state cannot be directly described in the static density functional
calculations. As a consequence, in a generated model unit cell, only a single hydrogen
bond is selected from the three virtual hydrogen bonds near each SeO$_4$ group which leads to 
a lowering of the symmetry of the system from the rhombohedral to monoclinic. In this case,
the generated structure can be considered as a static snapshot of 
the dynamically disordered superionic state. 
In the studies of the hydrogen transport mechanism we also considered a doubled unit cell, where
the initial orientations of the two hydrogen bonds in the cell are chosen randomly, each bond can 
be dynamically broked and created in a chain of intermediate configurations 
generated by the nudged elastic band (NEB) method \cite{qe,jonsson}.

In the studies of the transport mechanism, each
atomic configuration has been structurally relaxed. The relaxation involves the optimization
of the atomic positions by the minimization of the total energy
and forces acting on the atoms in the generated unit cell. For the structural optimization,
we use a Newton scheme described
in Ref.~\onlinecite{jonsson}. A comparison of different relaxed static configurations
allows to calculate the energy barriers for the proton migration.
With the obtained zero-temperature results, we also analyze the influence of temperature on the activation
energies of the proton transport.

\subsection{Modeling the hydrogen migration paths}\label{neb}

\begin{figure}[htbp]
\epsfxsize=8.5cm \centerline{\epsffile{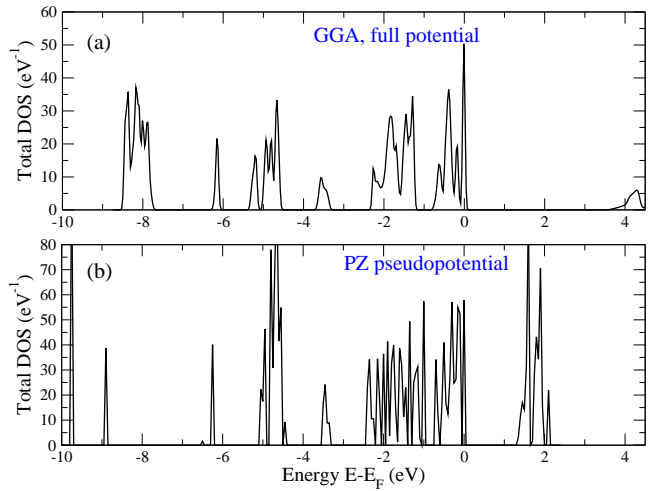}} \caption{Total density of states calculated
(a) in the GGA (Perdew, Burke, Ernzerhof, Ref.~\onlinecite{pbe}) using the full potential
approach of Ref.~\onlinecite{wien2k} and (b) in the LDA
using norm-conserving atomic pseudopotential approach with nonlinear core corrections
implemented by Perdew, Zunger, see Ref.~\onlinecite{perdew_zunger}. The zero
energies correspond to the Fermi levels.
} \label{fig3}
\end{figure}

To analyse in details the paths for the hydrogen migration in the relaxed surrounding,
we employed the nudged-elastic-band (NEB) method implemented in the
Quantum-Espresso (QE) Package of the DFT calculations with the use of plane-wave basis sets and
pseudopotentials \cite{qe,jonsson}. In these calculations, for the atomic
cores of Rb, H, Se and O we employ the Perdew-Zunger (PZ) norm-conserving
pseudopotentials with nonlinear core corrections \cite{perdew_zunger}.
Our choice of pseudopotentials is justified by the facts that all characteristic
features of the electronic structure calculated
with the use of the full potential Wien2k GGA code (Fig.~\ref{fig3}(a))
and obtained with the PZ-pseudopotentials (Fig.~\ref{fig3}(b)) are consistent in the energy window
($E_F-7.0$eV; $E_F+1$eV), where $E_F$ is the Fermi energy.
Specifically, both approaches allow to describe 
the characteristic DOS-peaks related to the O(2)-2p orbitals (the energy
range between $E_F$ and $E_F-5$~eV in Fig.~\ref{fig3}) and the DOS-peak
related to the 1s-orbital of H at the energy value about $-6.1$~eV (Fig.~\ref{fig3}(a)) which
corresponds to the peak at about $-6.2$~eV in Fig.~\ref{fig3}(b). In the GGA-approach, the energy gap 
between the highest filled and lowest empty electronic states is about 3.5~eV (Fig.~\ref{fig3}(a)) 
which is close to experimentally discussed gap about 4~eV~\cite{baranov3} in related compounds. The GGA gap
is larger then the gap about 1.3~eV obtained by the LDA-method (Fig.~\ref{fig3}(b)) due to the usual 
underestimation of the gap in the LDA calculations, which does not influence the results of the NEB studies
based on the calculations of the total energy. 
Also, we would like to note that
a detailed comparison of the electronic and optical properties of several compounds
like Rb$_2$O and Li$_2$O related to the
constituents of Rb$_3$H(SeO$_4$)$_2$ calculated in the LDA and GGA approaches shows a good
consistency with the available experimental results \cite{moakafi}.
As the low-temperature electronic
properties calculated by these two approaches
are consistent, we conclude that the PZ-pseudopotential approach
implemented in the QE-code
is sufficient to capture the physical features of the hydrogen migration in our system.

In contrast to the almost free motion of the hydrogen implemented in a wide range
of the models of hydrogen transport, in the present work we consider the migration of hydrogen
as a {\it cooperative process}, the feature which can be described by the NEB-approach \cite{qe,jonsson}.
Specifically, for each stage of the hydrogen transport, this process
involves a relaxation of the atomic positions
and of the distances between the different atoms. In our analysis, to better relax the positions of the atoms
neighbouring to H, we perform a doubling of the unit cell
which now includes four SeO$_4$ groups and two protons (see Fig.~\ref{fig8}(a) for details). For each
position of the hydrogen
in the unit cell, the coordinates of all neighbouring atoms were relaxed until the forces acting
on the atoms reached their minima.
In these calculations, we use the plane-wave cutoff
1020~eV and the energy cutoff for charge and potential given by 2040~eV.

In the NEB-approach, the relaxation of the atomic positions along the hydrogen migration path is
performed by the minimization of the total
energy of each intermediate configuration (image).
These images correspond to different positions of H on the migration path
and produced by the optimization of a specially
generated object functional (action) with the consequent minimization
of the spring forces perpendicular to the path. In our calculations,
the convergence criteria for the norm of the force orthogonal to the path is achieved at the values below 0.05eV/\AA.

\section{RESULTS AND DISCUSSION} \label{results}

\subsection{Vertex oxygens O(2)} \label{sect2b}

The experiments (see Ref.~\onlinecite{pawlowski}) show that in M$_3$H(XO$_4$)$_3$-crastals
the temperature of the superionic phase transition $T_S$
does not depend on the deuterization.
Moreover, the neutron diffraction results \cite{melzer}
do not distinguish two distinct hydrogen positions
on the bonds and justify a representation oh hydrogen atom in flat single-minimum potentials.
Consequently, in the present studies of the positioning of O(2),
the initial proton positions are fixed to the centers of hydrogen bonds. In Fig.~\ref{fig1}, such
a single-well approximation is identified by the centering of proton
between the oxygens O(2).

\begin{figure}[htbp]
\epsfxsize=8.5cm \centerline{\epsfclipon\epsffile{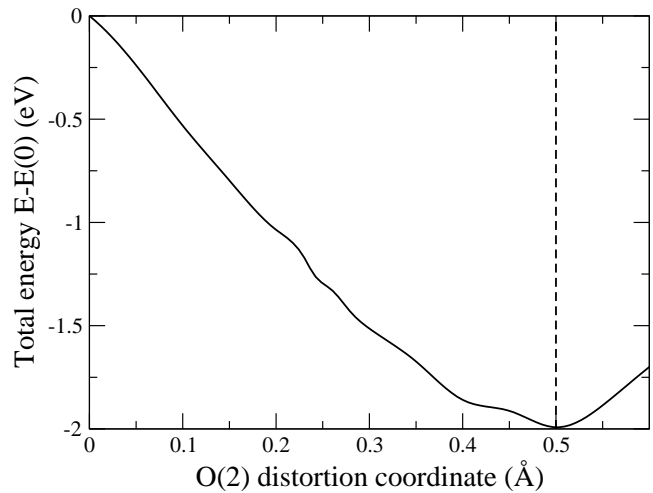}} \caption{
Total system energy as a function of the O(2) distortion parameter $\Delta$. Here $\Delta$
is the displacement of O(2) with respect to its high-symmetry position on the three-fold axis.
} \label{fig4}
\end{figure}

To study the modification of the total energy caused by the relaxation of the vertex
oxygens, the positions of O(2) on a selected hydrogen bond were shifted by a distance
$\Delta$, while the coordinates
of other atoms were kept in the undistorted high-symmetry state.
Fig.~\ref{fig4} demonstrates the variation of the total energy of the system $E(\Delta)-E(\Delta=0)$
with the increase of the O(2) displacement parameter $\Delta$. One can
see that the lowest value of $E(\Delta)$ corresponds to $\Delta=0.5$~\AA\,\, which implies a
stabilization of a short hydrogen bond O(2)-H-O(2)' of a length 2.54~\AA\,\, where the length
of the covalent bond O(2)-H equals 1.27~\AA. It is remarkable
that despite the central-point approximation for the proton within the hydrogen bonds,
the obtained length of the H-bond is typical for double-well H-bonds and
is consistent with the results
of x-ray and neutron scattering studies\cite{hilczer,baranov,melzer}, which becomes
clear from the comparison of different data with the present results presented in Table~\ref{tab1}. 
Consequently, our findings
clearly support the second concept of the oxygen-distorted network 
structure which is characterized
by the two types of dynamical disorder in the superionic phase:
(i) the statistical disorder of the hydrogens between the virtual hydrogen bonds in hexagonal planes
and (ii) the disorder of the vertex oxygens between three symmetry-related
positions around the three-fold axis indicated in Fig.~\ref{fig1}(b).

\begin{figure}[ht]
\epsfxsize=8.5cm \centerline{\epsffile{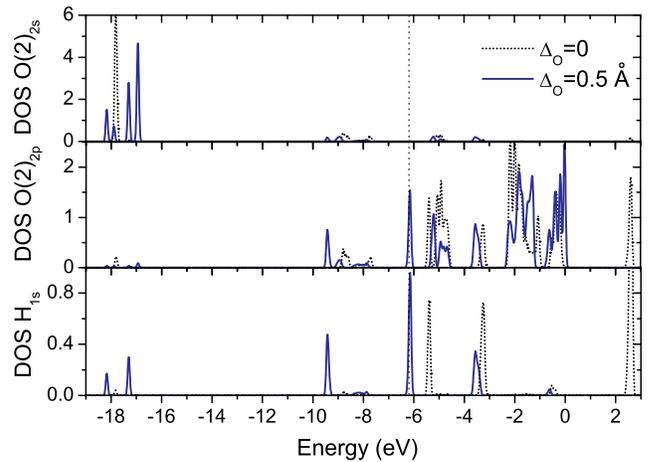}} \caption{
Projected density of states of the oxygens O(2) and hydrogens calculated by the GGA method. The cases $\Delta=0$
and $\Delta=0.5$~\AA\, correspond to the hydrogen bonds between SeO$_4$ groups (a) with high-symmetry
positions of O(2) on the three-fold axis and (b) with the displacements of O(2) from the three-fold
positions towards the proton. The zero energy corresponds to the Fermi level.
} \label{fig5}
\end{figure}

To demonstrate the influence of the displacements of O(2) on the electronic properties,
we have analyzed the projected atomic densities of electronic states for $\Delta=0$ and $\Delta=0.5$~\AA\, which are
presented in Fig.~\ref{fig5}.
The considered two types of the structural positioning of O(2) can be characterized by a substantial
electronic hybridization of H 1s orbitals and 2p orbitals of O(2). In Fig.~\ref{fig5}, the s-p
hybridization can be identified by the density peaks at the energies about $-3.5$~eV 
and in the range $(-5.5)$--$(-6)$~eV
below the Fermi energy (for $\Delta=0.5$~\AA, these peaks are marked by a dashed line). The
DOS peaks at about $-18$~eV imply also a strong hybridization between H 1s and O(2) 2s orbitals. The
displacements of O(2) from the high-symmetry positions lead to a shortening
of the O(2)-H distances and consequently to a significant increase of the s-p hybridization peaks at $-6$~eV.
Moreover, the O(2)-distortions result in the emergence of new strongly hybridized state at $-9.5$~eV.
In addition, the displacements of O(2) induce the
distortions of the SeO$_4$ tetrahedra which results in the lowering of the local symmetry
and in the split of high hybridized peaks around $-18$~eV (Fig.~\ref{fig5}, case $\Delta=0.5$~\AA).
In the considered model unit cell, instead of the virtual H-bonds we consider a number of selected
configurations with the fixed H-bonds near each SeO$_4$ group.
In a real statistically disordered system one may expect that the obtained
split of the hybridization peaks will be smoothed by the dynamical disorder.

\begin{figure}[ht]
\epsfxsize=8.5cm \centerline{\epsffile{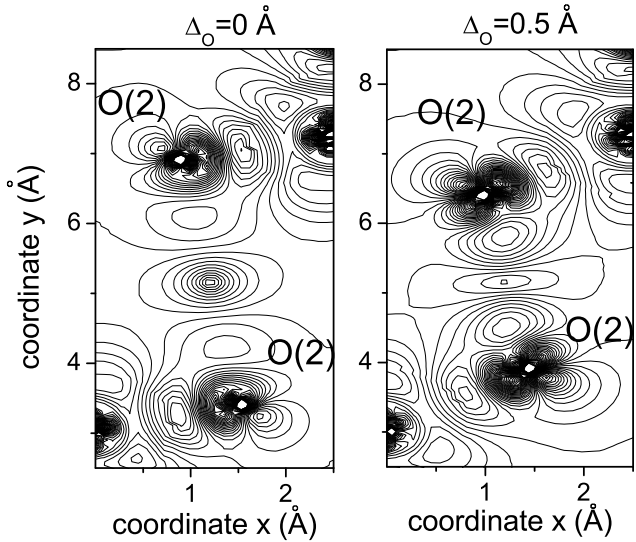}} \caption{
Electronic difference density contours (crystalline renormalized by the superimposed atomic densities)
for the hydrogen bonds O(2)-H-O(2). Here the oxygens O(2) are located on the three-fold
axis (left panel), and displaced from this axis towards the hydrogen on the bond (right panel).
} \label{fig6}
\end{figure}

The redistribution of the electronic charge density due to the distortions of O(2) causes
the accumulation of the electron charge on the obtained short 
hydrogen bonds, the effect observed in the contours
of the difference charge density in Fig.~\ref{fig6}.
Consequently, in the superionionic phase the formation of the strong
virtual hydrogen bonds should be also identified by a remarkable increase of 
the electron density between the O(2) ions, which is consistent with the experimentally
obtained electron density map of Fig~\ref{fig1}.

\begin{table}[b]
\caption{\label{tab1} Hydrogen bond length parameters (in \AA) in the superionic phase of Rb$_3$H(SeO$_4$)$_2$
determined from the experiments and calculated by the O(2)-relaxation in the full-potential GGA.
\\}

\begin{ruledtabular}
\begin{tabular}{llllllll}
GGA (O(2)-relaxation) & Ref.~\onlinecite{baranov} & Ref.~\onlinecite{melzer} & Ref.~\onlinecite{bohn}\\
\hline
2.54 & 2.51 & 2.594 & 2.58
\end{tabular}
\end{ruledtabular}
\end{table}

\subsection{Rotational stage in the full potential calculations \label{rot}}

\begin{figure}[htbp]
\epsfxsize=8.5cm \centerline{\epsfclipon\epsffile{fig7a.eps}} 
\epsfxsize=5.5cm \centerline{\epsfclipon\epsffile{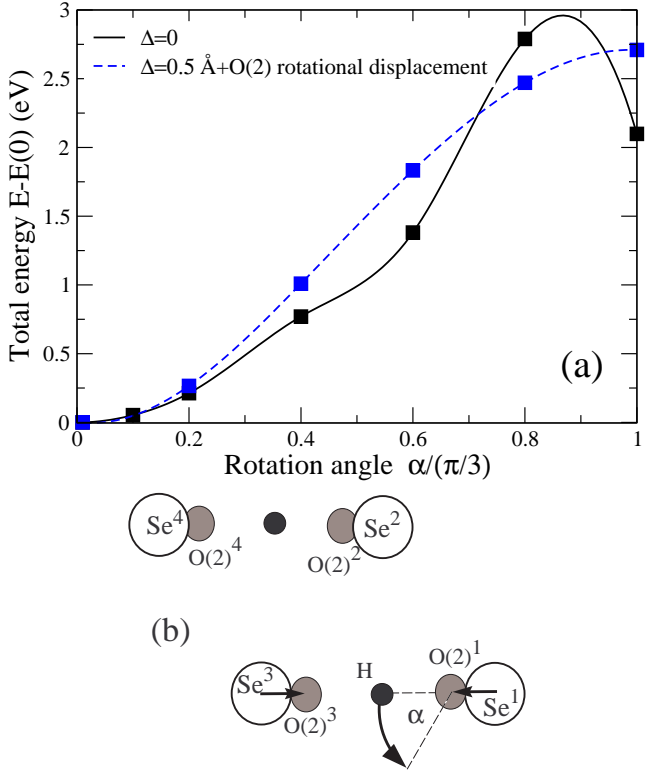}}
\caption{(a) Energy profiles for the rotation of
hydrogen around the group (SeO$_4$)$^1$. The case $\Delta=0$ corresponds to the oxygen O(2)$^1$ located
on the three-fold axis and fixed during the rotation of H. For $\Delta=0.5$~\AA, O(2)$^1$ is
displaced by a distance $\Delta$ and rotates jointly with the hydrogen. (b) Shematical view of the rotation stage
of the transport mechanism. The short arrows indicate the displacements of the oxygens O(2) from
the three-fold positions towards the proton, the large circles denote SeO$_3$ groups,
$\alpha$ is the rotation angle of O(2)-H covalent bond.
} \label{fig7}
\end{figure}

Based on the results obtained for the positions of the hydrogen-bonded O(2), it is possible to
study the mechanisms and paths of the hydrogen migration. In our analysis, the special focus
is put on the rotational stage of the Grotthuss transport mechanism. In its classical
description, this stage involves the rotational motion of the HSeO$_4$ group around the three-fold axis
determined by a rotation angle $\alpha$ (see Fig.~\ref{fig7}(b) where the rotation stage
is schematically shown), and the consequent formation of a
new hydrogen bond between the rotated group and the nearest SeO$_4$ tetrahedron approached by this rotation.

First, we analyze a possibility of the rotational step as an independent stage of the proton transport
mechanism. In this case, the hydrogen with the covalently bonded oxygen O(2) is allowed to rotate
around the nearest SeO$_3$-group while the positions of all the surrounding atomic groups are fixed.
In such a rotational stage, we have calculated the variation of the total energy of the system
due to the rotations. Similar to the studies
of the O(2) network in the previous section, we consider here 
two possible structural concepts for the rotational motion.
In the first high-symmetry case, the oxygen O(2)$^1$ of the H(SeO$_4$)$^1$ group 
is fixed on the three-fold axis ($\Delta=0$),
whereas in the second configuration the oxygen O(2)$^1$ is relaxed (displaced) by a distance $\Delta=0.5$~\AA\,
from the three-fold axis and rotates jointly with the hydrogen by the angle $\alpha$
(the rotation of O(2)$^1$-H covalent bond).
The resulting dependences of the total energy on $\alpha$ are presented in Fig.~\ref{fig7}.

It is remarkable that both, high-symmetry and O(2)$^1$-H-rotated configurations 
are characterized by high-energy barriers (about $2.6-3$~eV)
between the initial position of proton and the transition state (TS) (Fig.~\ref{fig7}(a)) which corresponds
to the intermediate value of the rotation angle $\alpha$ about $\pi/3$. The high energy
barriers {\it absolutely exclude any independent motion of proton in the conducting system}, a
concept which is widely applied for the free electrons in metals. We also note the obtained values
for the energy barriers are af the same order as the rotational barrier $1.54$~eV calculated in Ref.~\onlinecite{ke} 
for the system CsHSeO$_4$ which prompts on the common physical aspects of the transport mechanisms 
in these classes of proton conductors.

Despite the high energy of the TS state, the dashed energy profile in Fig.~\ref{fig7} demonstrate
a possibility for a partial rotation of the covalent bond O(2)$^1$-H by $\alpha=20-25~\deg$ which has
considerably low barriers of about $0.3-0.7$~eV. In the superionic state,
such a partial rotation can be also accompanied by the displacements of the oxygen O(2)$^2$
of a neighbouring group (SeO$_4$)$^2$ towards the hydrogen which would lower the energy barrier and
could result in the formation of the new hydrogen bond O(2)$^1$-H-O(2)$^2$ with the
consequent proton transfer to the group (SeO$_4$)$^2$. As a conclusion,
the rotational motion of protons should also involve the relaxation of the oxygens O(2) of
more distant SeO$_4$ groups.

\subsection{Hydrogen migration as a cooperative process}

\begin{figure}[ht]
\epsfxsize=6.5cm \centerline{\epsffile{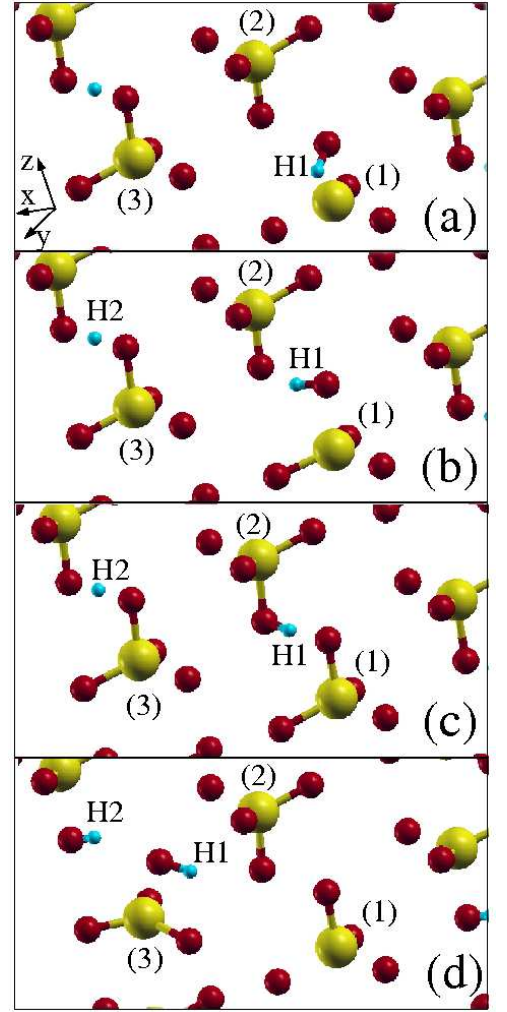}}
\caption{Snapshots demonstrating hydrogen migration between the neighbouring groups (SeO$_4$)$^1$
and (SeO$_4$)$^3$.
Here Se, O and H atoms are represented by yellow, red and blue circles.
Sections (a) and (b) show configurations with rotated (SeO$_4$)$^1$; (c) resulted from the transfer
from (1) to (2) and rotation around (SeO$_4$)$^2$, configuration (d) is obtained after
the transfer of hydrogen between two tetrahedra (SeO$_4$)$^2$ and (SeO$_4$)$^3$.
} \label{fig8}
\end{figure}

\begin{figure}[ht]
\epsfxsize=8.2cm \centerline{\epsfclipon \epsffile{fig9ab.eps}}
\epsfxsize=5.0cm \centerline{\epsffile{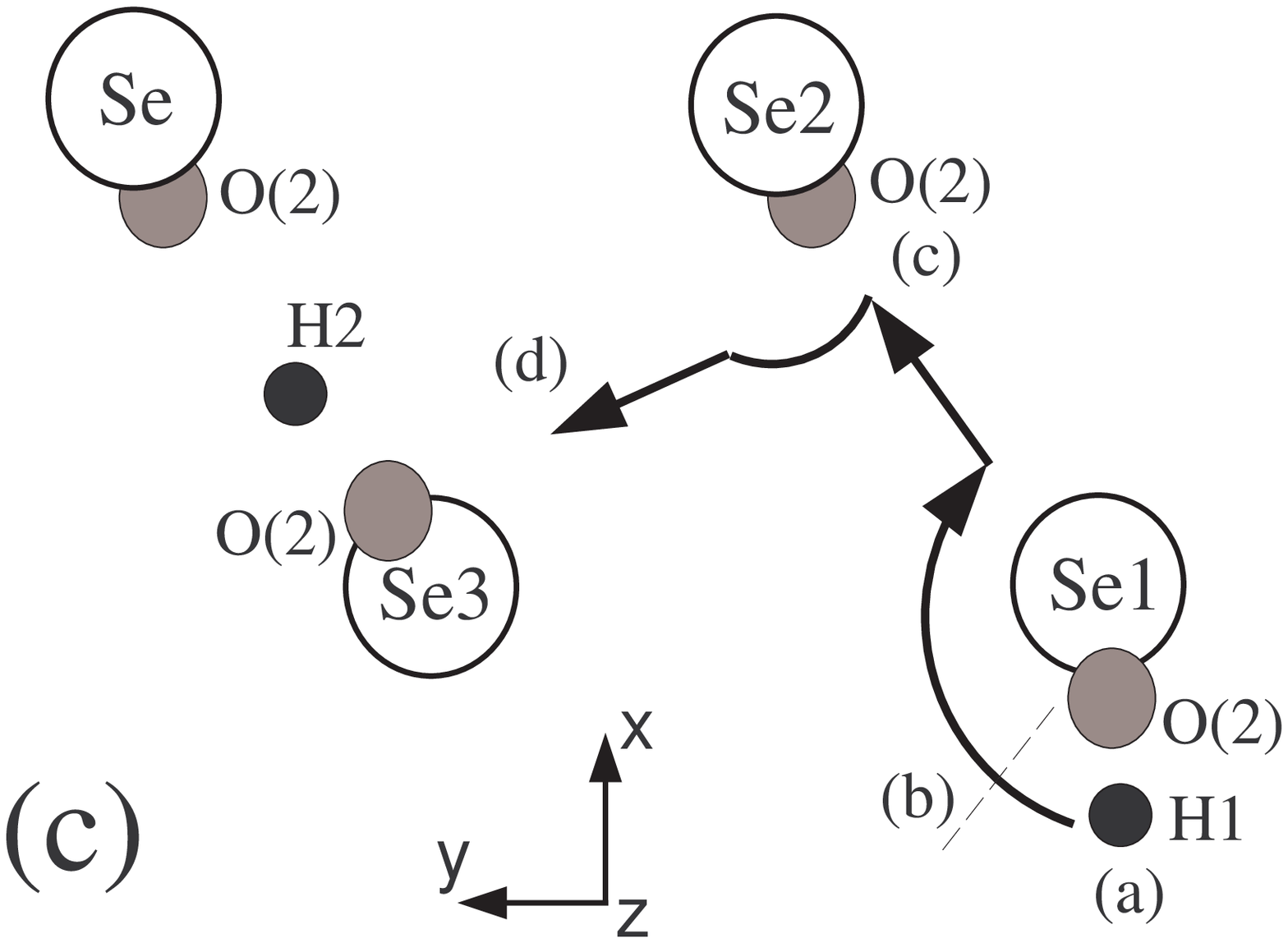}}
\caption{Energy profiles for a migration of
hydrogen between the tetrahedra (SeO$_4$)$^1$ and (SeO$_4$)$^2$. Here the part (a) corresponds to
the rotation of the covalent bond H1-O(2)
around the group (SeO$_4$)$^1$ and the part (b) corresponds to the transfer of H1 to the group (SeO$_4$)$^2$.
(c) Schematic view of the migration path of H1 in the projection on the hexagonal (001)-plane
where the small symbols (a)--(d) correspond
to the images in Fig.~\ref{fig8}.
} \label{fig9}
\end{figure}

In the studies of hydrogen migration in a fully relaxed surrounding, we employ the NEB-mathod 
discussed in section~\ref{neb}.
Fig.~\ref{fig8} shows the snapshots of the migration of the hydrogen H1 between the complexes 
(SeO$_4$)$^1$-(SeO$_4$)$^2$ and (SeO$_4$)$^2$-(SeO$_4$)$^3$ calculated by the NEB method.
The obtained proton migration path can be described via several steps: 

(1) the O(2)$^1$-H1 covalent bond is rotated around SeO$_3^1$-group which corresponds
to the variation of the rotation angle $\alpha$ from its initial value $\pi/3$ 
to the value $\alpha=0$ (part (a) in Fig.~\ref{fig8} is schematically shown in Fig.~\ref{fig9}(c)). 
We note that this rotation fully corresponds to the rotational motion of H studied in the previous section~\ref{rot}
within the full potential approach, whereas the rotation in section~\ref{rot} 
is performed in the opposite direction shown in Fig.~\ref{fig6}(b) with
the angle $\alpha$ changing from 0 to $\pi/3$. The comparison of the corresponding energy profiles
calculated by the NEB-method (Fig.~\ref{fig9}(a), H1
coordinate $r$ in the range $0 \le r\le 0.8$\AA) and in the full-potential approach 
(Fig.~\ref{fig6}) shows that despite the similar parabolic form, 
the energy barrier in the QE-case $\Delta E(QE)=E(0)-E(r=0.8)$ is 
about 1.0~eV which is much lower then the full-potential GGA-value obtained
from the difference $\Delta E(FP)=E(\alpha=\pi/3)-E(\alpha=0)=2.72$~eV. 
This comparison clearly demonstrates the key role of 
the structural relaxation in the NEB-approach for the lowering the energy barriers 
for the proton transport.

\begin{figure}[ht]
\epsfxsize=8.2cm \centerline{\epsfclipon \epsffile{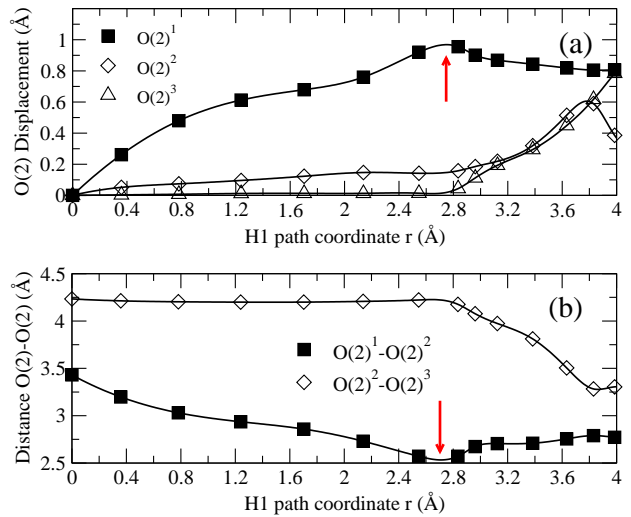}}
\caption{(a) Displacements of the vertex oxygens O(2) and (b) the change of the
distance between the nearest O(2) along the migration path of H1. The red arrows identify
the transition state of the proton between the hydrogen atoms O(2)$^1$ and O(2)$^2$.
} \label{fig10}
\end{figure}

In the further rotation of O(2)$^1$-H1 the H1 coordinate $r$ 
changes from $r=0.8$ to $r=2$\AA\, which leads to a small increase of the total energy by about 0.2eV
in Fig.~\ref{fig9}(a). In Fig.~\ref{fig9}(a), the minimum of the total energy at $r=0.8$\AA\, 
corresponds to the nearly zigzag alignment
of the O(2)-H-dimers (configuration presented in part (b) of Fig.~\ref{fig8}), similarly to the 
low-temperature ordered phase III stabilized below the superionic transition
temperature, see Refs.~\onlinecite{baranov, stasyuk,pavlenko3} for the detailed description of the structure of this 
phase. The rotation part in the range $0.8 \le r\le 2$~\AA\, corresponds to the rotation stage
of the Grotthuss transport mechanism.

In Fig.~\ref{fig9}(a), the energy barrier for the hydrogen rotation step approaches 0.2~eV.
Here the rotation barrier is estimated from the energy dependence between $r=0.8$ (proton position indicated by
the dashed line in Fig.~\ref{fig9}(c)) and $r=2$\AA.

(2) The hydrogen is transferred from the oxygen ion O(2)$^1$ (part (b) in
Fig.~\ref{fig8}) to the oxygen O(2)$^2$ (part (c) in Fig.~\ref{fig8}). 
Fig.~\ref{fig10} presents
the displacements of the oxygens O(2) and the changes of the O(2)-O(2) distances during the migration
of the hydrogen derived from the DFT(NEB)-calculations. 
The proton transfer stage occurs in the range ($2<r< 3$) of the 
proton coordinate. It is remarkable
that this range of $r$ is characterized by significant displacements of O(2)$^1$ and O(2)$^2$ by 0.2-0.25~\AA\, towards 
the hydrogen and by the corresponding strong decrease of the distance
between O(2)$^1$ and O(2)$^2$ from its inital value about 3\AA\, to the value 2.51\AA, which
is very close to the results of the full-potential GGA studies (see Table~\ref{tab1}). In Fig.~\ref{fig10}, the 
occurrence of the transfer of the hydrogen between O(2)$^1$ and O(2)$^2$ is identified by the minimal
O(2)-O(2) distance and maximal displacement of O(2)$^1$ which is indicated by the red arrows in the plots.
Table~\ref{tab2} presents the maximal ($d_{\rm OO}^{\rm max}$) and minimal ($d_{\rm OO}^{\rm min}$) lengths of the 
hydrogen bond during the transfer step calculated
in the NEB approach which are compared to the corresponding experimentally obtained values. 
Despite the good agreement with the experimental maximal bond length, the calculated
value of $d_{\rm OO}^{\rm min}=2.51$~\AA\, is in coincidence with the estimates of Ref.~\onlinecite{baranov} for the
superionic Rb$_3$H(SeO$_4$)$_2$ but is 
larger then the value $2.41$~\AA\, discussed in Ref.~~\onlinecite{hilczer} for the similar system 
(NH$_4$)$_3$H(SeO$_4$)$_2$. The difference is possibly related to the influence of the dynamics of 
NH$_4^+$ groups in the superionic state which leads to additional softening of the structure and to the 
consequent increase of the O(2) displacements \cite{hilczer}.

\begin{table}[b]
\caption{\label{tab2} Hydrogen bond parameters (minimal and maximal O(2)-O(2) distances $d_{\rm OO}^{\rm min}$ and
$d_{\rm OO}^{\rm max}$ in \AA), protonic polaron transport parameter $\nabla V$ (eV/\AA) and the transport 
activation energy $E_a$(eV) in the superionic phase of Rb$_3$H(SeO$_4$)$_2$
calculated by the NEB-method and estimated experimentally.
\\}

\begin{ruledtabular}
\begin{tabular}{llllllll}
method & $d_{\rm OO}^{\rm min}$ & $d_{\rm OO}^{\rm max}$ & $\nabla V=\frac{\partial E}{\partial 
\Delta_{\rm O}}$ & $E_a$ \\
\hline
NEB (full relaxation) & 2.51 & 2.7 & 2.55 & 0.33 \\
Refs.~\onlinecite{hilczer,pawlowski} & 2.4 & 2.7 & - & 0.37  \\
Ref.~\onlinecite{baranov} & 2.51 & 2.67 & - & 0.49
\end{tabular}
\end{ruledtabular}
\end{table}

(3) The O(2)$^2$-H1 covalent bond rotates and the hydrogen H1 moves towards the group (SeO$_4$)$^3$.  The process
of the O(2)$^2$-H1 rotation in the group (SeO$_4$)$^2$ is accompanied by the 
elongation of the covalent bond between the oxygen O(2)$^3$
and another proton H2 (shown in the left corner of part (c)) and by a strong displacement of the oxygen O(2)$^3$ 
towards the hydrogen H1 (part (d) in Fig.~\ref{fig8}). 
The displacement of the oxygen O(2)$^3$ leads to a decrease of the 
distance between O(2)$^2$ and O(2)$^3$ from its initial value 4.2~\AA\, to the value 3.22~\AA\,(Fig.~\ref{fig10}(b)) 
and is essential for the successful hydrogen migration to the group (SeO$_4$)$^3$. 
At the transfer of H1 between O(2)$^2$ and O(2)$^3$, the
lengths of the bonds Se-O(2)$^3$ (vertex oxygen) and Se-O(1)$^3$ (basis oxygen) in the group (SeO)$^3$ are 
also changed by about 0.08--0.16\AA. After the proton is 
transferred from O(2)$^2$ to O(2)$^3$, the length of the Se-O(2)$^2$ bond is
returned to its previus value 1.76\AA. 

\begin{figure}[ht]
\epsfxsize=8.5cm \centerline{\epsfclipon \epsffile{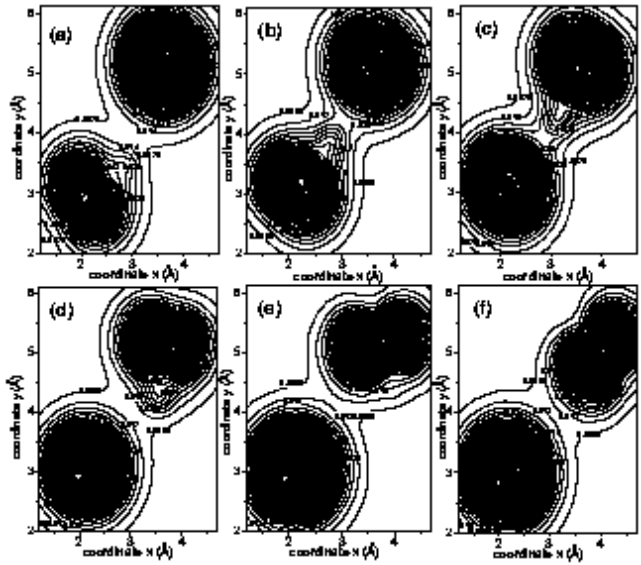}}
\caption{Electronic density contours showing the redistribution of the valence charge
density of the hydrogen bond O(2)$^1$-H-O(2)$^2$ during the intrabond proton transfer 
(a)-(d) and during the
the rotational motion of H covalently bonded to O(2)$^2$ (d)-(f). The valence charge has been obtained by the
integration of DOS in the energy window ($E_F-4$~eV; $E_F$) where $E_F$ is the Fermi level.
} \label{fig11}
\end{figure}

Fig.~\ref{fig11} demonstrates the redistribution of the valence charge density during the migration of the hydrogen.
The intrabond proton transfer step between the vertex oxygens O(2)$^1$ and O(2)$^2$ 
corresponds to the contours (a)-(d) of the plot. One can see that the hydrogen transfer 
has a character of a continuous redistribution of the electronic charge density between the two different
 positions near O(2)$^1$
and near O(2)$^2$ detected in the
plots (b) and (c). This finding allows to improve the conclusions of the neutron scattering studies 
in Ref.~\onlinecite{melzer} about the absence of the two distinct proton positions within a hydrogen bridge
in the conducting phase of Rb$_3$H(SeO$_4$)$_2$ which can be explained by the limited resolution of the 
experiments. Furthermore, in the rotational step (plots (d)-(f)), the rotational displacements of the oxygen
O(2)$^2$ follow the proton rotation and lead to significant modifications of the charge 
density contours near O(2)$^2$ which obtain a distinct dumbbell character. 

As the consequence of our findings, we propose to consider 
the obtained in our studies {\it significant displacements of the oxygens O(2)} 
during the migration of the hydrogen as {\it an additional feature of the proton transport mechanism}. 
It is also remarkable that 
the well established in the literature standard Grotthuss proton transport mechanism contains two main 
(rotational and translational) stages and does not involve the 
oxygens displacements as a decisive factor in the proton migration 
process, which is in contrast to the results of the present work. 

To see the role of O(2) displacements in the establishing the hydrogen transport,
we examine more closely the change of the total system energy along the proton 
migration path presented in Fig.~\ref{fig9}.

Fig.~\ref{fig9} shows that the energy barrier for the hydrogen rotation step approaches 0.2~eV and is substantially lower
than the barrier for the proton transfer between neighbouring groups which is about 0.6~eV in Fig.~\ref{fig9}(b). 
The relation between the obtained rotation and translation barriers is consistent with the conclusion about
the decisive role of the intrabond proton transfer in the determining the activation energy
for proton conductivity reported in Refs.~\onlinecite{lechner,hilczer}. 

In M$_3$H(XO$_4$)$_2$ systems, the 
dominant role of the intrabond translational step in the formation of the transport activation energy is in
contrast to the quasi-one-dimensional proton conductors MHXO$_4$, 
where the proton conduction process is determined by the 
rotational energy barriers. This fact is also supported by the recent DFT studies of CsHSO$_4$~\cite{ke} where the 
intrabond proton barrier 0.16~eV was found to be substantially lower than the rotational barrier 0.52~eV, and is 
also consistent with the ab-initio molecular dynamics calculations of the related systems ~\cite{lee} which show 
strong fluctuations
of the hydrogen bond lengths between 2.4 and 2.8~\AA\, during the proton migration and fast almost barrierless proton 
transfer on the hydrogen bonds characterized by the extremely short lengths about 2.4~\AA. The fast intrabond
proton transfer in the quasi-one-dimentional proton conductors is similar to the proton transport properties
in water mixtures where the recent ab-initio molecular dynamics simulations demonstrate the existence of
two different time scales and identify the faster process with the intrabond proton transfer, while the slower time 
scale corresponds to local rearrangements of the hydrogen-bond network ~\cite{morrone}. In this context, the 
unique two-dimensional character of the transport mechanism in the M$_3$H(XO$_4$)$_2$ introduces qualitative 
changes in the translational and rotational energy barriers with a consequent importance of the intrabond transfer
in the determination of the transport energy barriers.

The change of the total energy profile along the transfer path in Fig.~\ref{fig9}(b) can be directly compared 
with the corresponding displacements of hydrogen bonded O(2) in Fig.~\ref{fig10}(a). From this comparison,
we obtain that the increase of the total energy up to $\Delta_E=0.55$~eV is accompanied by the O(2)-displacements $\Delta_{\rm 
O}=0.25$~\AA. In the phenomenological description of the proton transport\cite{pavlenko}, the 
anti-phase displacements of O(2) induced by the hydrogen-bonded proton have been classified in terms of the 
``proton-polaronic'' effect where the protonic polaron is formed due to the interactions of proton with the anti-phase
stretching displacements of the nearest O(2)-oxygens and 
the energy for the formation of the protonic polaron $E_0=(\hbar\nabla V)^2/2M(\omega_0)^2$
depends on the energetic parameter $\nabla V=\partial E/\partial \Delta_{\rm O}$ (here $M$ is the oxygen mass) and on the 
characteristic frequency $\omega_0$ of 
of the O(2)-stretching vibration mode. From the {\it ab-initio}-profiles shown in Fig.~\ref{fig9}(b) and Fig.~\ref{fig10}(a),
one can derive the value of $\nabla V=2.55$~eV\AA$^{-1}$ which appears to be very close to the value 2.4~eV\AA$^{-1}$ used
in the phenomenological calculations in Ref.~\onlinecite{pavlenko}. Using the value $\omega_0\approx 850$~cm$^{-1}$ 
which is in the range of the characteristic frequencies of the HSeO4 stretching vibrations of Rb$_3$H(SeO$_4$)$_2$
reported in Ref.~\onlinecite{pawlowski2}, we obtain the estimate for the protonic polaron binding energy: $E_0=0.2$~eV.
With this value, using a high-temperature estimate $E_a\approx 5E_0/3$ obtained in the weak proton interaction 
limit from Ref.\onlinecite{pavlenko}, we find 
the activation energy for the proton transport $E_a=0.33$~eV. Table~\ref{tab2} compares the activation energies for
the proton conductivity derived on the basis of the present DFT-calculations and reported in different experimental measurements. 
The obtained here value 0.33~eV is very close to the experiment which confirms a central role of the O(2)-distortions
for the proton migration mechanism.

It is also worth noting that the snapshot in Fig.~\ref{fig8}(d) corresponds in fact to the 
two protons located near the same ionic group (SeO$_4$)$^3$, a configuration similar 
to the high-energy Bjerrum (rotational) 
defect \cite{stasyuk3}. The energy barrier for the creation of such type of defects is 
calculated from the obtained energy profile 
and is about 1.5eV which is twice as high as the estimated translation barrier and which supports negligibly small
probability for the creation of the Bjerrum defects in this family of proton conductors.

\section{Conclusions} \label{conclusions}

We have performed the density functional studies of the electronic properties and the mechanisms of hydrogen
migrations in the system Rb$_3$H(SeO$_4$)$_2$ which belongs to the technologically promising class of  
proton conductors of M$_3$H(XO$_4$)$_2$ crystal family (M=Rb,Cs, NH$_4$; X=S,Se).
The results of the 
electronic structure calculations show a central role of the lattice dynamics in the
process of the proton migration.
Our findings verify several experimental conclusions obtained in the
previous x-ray and NMR experiments with Rb$_3$H(SeO$_4$)$_2$, which allow to clalify the
positioning
of the vertex oxygens O(2) during the proton migration and to examine the proton positions
on the hydrogen bond at the intrabond proton transfer.
The principal conclusion obtained in this work is that the free rotational motion of protons without 
the corresponding complex distortions of SeO$_4$ groups cannot be considered as an independent 
part of the Grotthuss transport mechanism. In contrast to the almost free proton motion implemented in a wide range of the 
models of proton transport, in our work we have considered the migration of hydrogen as a cooperative process which 
involves a relaxation of the atomic positions and of the distances between the different 
ionic groups XO$_4$. 
We have shown that the migration of the hydrogens is connected with the significant displacement
of the vertex oxygens O(2). These displacements play a decisive role in the formation of the protonic
polaron and in the consequent decrease of the activation energy for the proton conductivity, 
a concept introduced in the previous phenomenological description of the proton transport. The present DFT results
allow to obtain the first-principle
estimates of the hydrogen-bond lengths and energetic parameters involved in the existing 
phenomenological modeling of the proton transport and to prove the validity of the protonic polaron model.
Due to their key role in the formation of new hydrogen bonds and in the 
establishing the sucessfull intrabond proton transfer, the oxygen displacements 
should be consideted as essential feature of the
obtained revised mechanism of the hydrogen transport. 
The developed approach can be also implemented for the modelling of a wide range of the
proton conductors which in our view substantially increases the degree of scientific and 
possible technological applications of the obtained results.

\section*{Acknowledgements}
This work was supported through the grants of computer time
from the Poznan Supercomputing Center and from the Ukrainian Academic Grid.


\section*{References}\label{refs}

\begin{thebibliography}{10}

\bibitem{norby} T.~Norby, Nature {\bf 410}, 877(2001).

\bibitem{haile} S.M.~Haile, D.A.~Boysen, C.R.I.~Chisholm, and R.B.~Merle, Nature {\bf 410}, 877(2001).

\bibitem{nature1} J.~Tollefson, Nature {\bf 464}, 1262(2010); {\it ibid.} Nature {\bf 460}, 442(2009).

\bibitem{dft1} J.S.~Hummelshoej et al., J.~Chem.~Phys. {\bf 131}, 014101(2009).

\bibitem{merinov} B.~Merinov and W.~Goddard, J.~Chem.~Phys. {\bf 130}, 194707(2009).

\bibitem{bjorketun} M.E.~Bj\"orketun, P.G.~Sundell, and G.~Wahnstr\"om, Phys.~Rev.~B {\bf 76}, 054307(2007).

\bibitem{zhang} Q.~Zhang, G.~Wahnstr\"om, M.E.~Bj\"orketun, S.~Gao, and E.~Wang, Phys.~Rev.~Lett. {\bf 101}, 
215902 (2008).

\bibitem{belushkin} A.V.~Belushkin, C.J.~Carlile, L.A.~Shuvalov, Ferroelectrics {\bf 167}, 83(1995).

\bibitem{lechner} R.E.~Lechner, Ferroelectrics {\bf 167}, 83(1995).

\bibitem{yamada} Y.~Yamada, Ferroelectrics {\bf 170}, 23(1995).

\bibitem{bohn} A.~Bohn, R.~Melzer, R.~Sonntag, R.E.~Lechner, G.~Schuk, and K.~Lange,
Sol.~State.~Ionics {\bf 77}, 111(1995).

\bibitem{merinov1} B.V.~Merinov, N.B.~Bolotina, A.I.~Baranov, and L.A.~Shuvalov, Sov.~Phys.~Crystallogr. {\bf 33}, 1387(1988).

\bibitem{baran2} B.V.~Merinov, A.I.~Baranov, and L.A.~Shuvalov, Sov.~Phys.~Crystallogr. {\bf 35}, 355(1990).

\bibitem{merinov2} B.V.~Merinov, M.Yu.~Antipin, A.I.~Baranov, A.M.~Tregubchenko, L.A.~Shuvalov, and Yu.T.~Struchko,
Sov.~Phys.~Crystallogr. {\bf 36}, 872(1991).

\bibitem{baranov} A.I.~Baranov, I.P.~Makarova, L.A.~Muradyan, A.V.~Tregubchenko, L.A.~Shuvalov, V.I.~Simonov,
Sov.~Phys.~Crystallogr. {\bf 32}, 400(1987).

\bibitem{dolinsek} J.~Dolinsek, U.~Mikas, J.E.~Javorsek, G.~Lahajnar, and R.~Blinc, L.F.~Kirpichnikova,
Phys.~Rev.~B {\bf 58}, 8445(1998).

\bibitem{hilczer} A.~Pietraszko, B.~Hilczer, and A.~Pawlowski, Sol.~State~Ionics
{\bf 119}, 281(1999).  

\bibitem{plakida} N.M.~Plakida, W.~Salejda, Phys.~Stat.~Sol(b) {\bf 148} 473(1988).

\bibitem{kamimura} T.~Ito, H.~Kamimura, J.~Phys.~Soc.~Jap. {\bf 67}, 1999(1998).

\bibitem{stasyuk} I.V.~Stasyuk, N.~Pavlenko, B.~Hilczer, Phase~Trans. {\bf 62}, 135(1977).

\bibitem{pavlenko} N.I.~Pavlenko, J.Phys.:Cond.~Matter {\bf 11}, 5099(1999).

\bibitem{pavlenko2} N.I.~Pavlenko, I.V.~Stasyuk, J.~Chem.~Phys. {\bf 114}, 4607(2001).

\bibitem{chisholm} C.R.I.~Chisholm, Y.H.~Jang, S.M.~Haile, and W.A.~Goddard~III, Phys.~Rev.~B {\bf 72}, 134103(2005).

\bibitem{ke} X.~Ke and I.~Tanaka, Phys.~Rev.~B {\bf 69}, 165114(2004).

\bibitem{wien2k} P.~Blaha {\it et al.}, {\it WIEN2K},
{\it An Augmented Plane Wave + Local Orbitals Program for Calculating Crystal Properties},
ISBN 3-9501031-1-2 (TU Wien, Austria, 2001).

\bibitem{pbe} J.P.~Perdew, S.~Burke, and M.~Ernzerhof, Phys.~Rev.~Lett. {\bf 77}, 3865(1996).

\bibitem{perdew_zunger} J.P.~Perdew and A.~Zunger, Phys.~Rev.~B {\bf 23}, 5048(1981).

\bibitem{qe} P.~Giannozzi et al., J.~Phys.~Cond.~Matter {\bf 21}, 395502(2009).

\bibitem{pawlowski} A.~Pawlowski, Cz.~Pawlaczyk, and B.~Hilczer, Solid State Ion. {\bf 44}, 17(1990).

\bibitem{melzer} E.~Melzer, T.~Wessels, and M.~Reehuis, Sol.~State~Ionics {\bf 92}, 119(1996).

\bibitem{jonsson} H.~Jonsson, G.~Mills, and K.W.~Jacobsen, {\it Nudged Elastic Band
Method for Finding Minimum Energy Paths of Transitions in Classical and Quantum
Dynamics in Condensed Phase Simulations} in {\it Classical and Quantum
Dynamics in Condensed Phase Simulations}, ed. by B.J.~Berne, G.~Ciccoti, and D.F.~Coker 
(Singapore: World Scientific, 1998).

\bibitem{baranov3} A.I.~Baranov, L.A.~Shuvalov, and N.M.Shchagina, JETP Letters, {\bf 36} 459(1983).

\bibitem{moakafi} M.~Moakafi, R.~Khenata, A.~Bouhemadou, H.~Khachai, B.~Amrani, D.~Rached,
and M.~Rerat, Eur.~Phys.~J.~B {\bf 64}, 35(2008).

\bibitem{pavlenko3} N.~Pavlenko, Phys.~Rev.~B {\bf 61} 4988(2000).

\bibitem{lee} H.-S.~Lee and M.E.~Tuckerman, J.~Phys.~Chem.~C {\bf 112}, 9917(2008).

\bibitem{morrone} J.A.~Morrone, K.E.Haslinger, and M.E.~Tuckerman, J.~Phys.~Chem.~B {\bf 110}, 3712(2006).

\bibitem{pawlowski2} A.~Pawlowski and M.~Polomska, Sol.~State~Ionics {\bf 176}, 2045(2005).

\bibitem{stasyuk3} I.V.~Stasyuk, O.L.Ivankiv, and N.I.~Pavlenko, J.~Phys.~Stud. {\bf 3}, 419(1997).

\end{thebibliography}
\end{document}